\documentclass[aps,prb,reprint,showpacs,superscriptaddress]{revtex4-1}

\usepackage{graphics}
\usepackage{bm}
\begin{document}
\preprint{In preparation, to be submitted to Physical Review}
\title{ Steadfast perpendicular exchange coupling in an ultrathin CoO/PtFe double-layer: strain and spin orientation }
\author{Anne D. Lamirand}
\affiliation{Institut N\'eel, CNRS and UJF, BP166, 38042 Grenoble, France}
\author{M\'arcio M. Soares}
\affiliation{Institut N\'eel, CNRS and UJF, BP166, 38042 Grenoble, France}
\altaffiliation{present address: European Synchrotron Radiation Facility - ESRF, Grenoble, France}
\author{Aline Y. Ramos}
\affiliation{Institut N\'eel, CNRS and UJF, BP166, 38042 Grenoble, France}
\author{H\'elio C. N. Tolentino}
\email[]{helio.tolentino@grenoble.cnrs.fr}
\affiliation{Institut N\'eel, CNRS and UJF, BP166, 38042 Grenoble, France}
\author{Maurizio De Santis}
\affiliation{Institut N\'eel, CNRS and UJF, BP166, 38042 Grenoble, France}
\author{Julio C. Cezar}
\affiliation{Laborat\'orio Nacional de Luz S\'incrotron-LNLS, CP 6192, 13083-970 Campinas, Brazil}
\author{Abner de Siervo}
\affiliation{Instituto de F\'isica Gleb Wataghin, Universidade Estadual  de Campinas-UNICAMP, 13083-970 Campinas, Brazil}
\author{Matthieu Jamet}
\affiliation{Institut Nanosciences et Cryog\'enie-INAC, CEA, 38042 Grenoble, France}
\date{\today}

\begin{abstract}
We report on the exchange coupling and magnetic properties of a strained ultrathin CoO/PtFe double-layer with perpendicular magnetic anisotropy. The cobalt oxide growth by reactive molecular beam epitaxy on a Pt-terminated PtFe/Pt(001) surface gives rise to a hexagonal surface and a  monoclinic distorted CoO 3nm film at room temperature. This distorted ultrathin CoO layer couples with the PtFe(001) layer establishing a robust perpendicular exchange bias shift. Soft  x-ray absorption spectroscopy  provides a full  description of the spin orientations in the CoO/PtFe double-layer. The  exchange bias shift  is preserved up to the N\'eel antiferromagnetic ordering temperature of ${T}_{N}$=293 K. This  unique example of selfsame value for blocking and ordering temperatures, yet  identical to the bulk ordering temperature, is likely related to the original strain induced distortion and strengthened interaction between the two well-ordered spin  layers. 
\end{abstract}

\maketitle

The conception and optimization of tuned devices for spintronic applications \cite{Chappert2007NMat} stir up a great interest in the exchange coupling between antiferromagnetic (AFM) and ferromagnetic (FM) layered materials \cite{He2010NMat, Park2011NMat, Shiratsuchi2012PRL,Berkowitz1999JMMM} and, particularly, in the unidirectional anisotropy effect known as exchange bias (EB) \cite{Meiklejohn1956PR}. The AFM/FM exchange coupling relies on a variety of microscopic and atomic parameters, as crystallographic order, surface morphology,  strain effects, spin orientation and competing anisotropies  \cite{Berkowitz1999JMMM}. The EB effect is largely used to pin the FM magnetization along one orientation in a spin valve or magnetic tunnel junction  \cite{Chappert2007NMat,He2010NMat, Park2011NMat, Shiratsuchi2012PRL,Berkowitz1999JMMM}.  It also provides greatest  opportunities to explore phenomena interlinking the spin and charge degrees of freedom \cite{He2010NMat} and, more recently, to  control the electronic transport in a tunneling anisotropic magnetoresistance device \cite{Park2011NMat}. In the latter the tunneling resistance is strongly affected by the orientation of the magnetic moments in the AFM layer, which can be partially rotated by the exchange coupling with the FM layer. Devices showing EB perpendicular to the layered surface are especially promising for low power consumption and ultrafast circuits, as well as for high-performance memories \cite{Shiratsuchi2012PRL, Mangin2006NMat}. 

\begin{figure}
\resizebox{1\columnwidth}{!}{\includegraphics{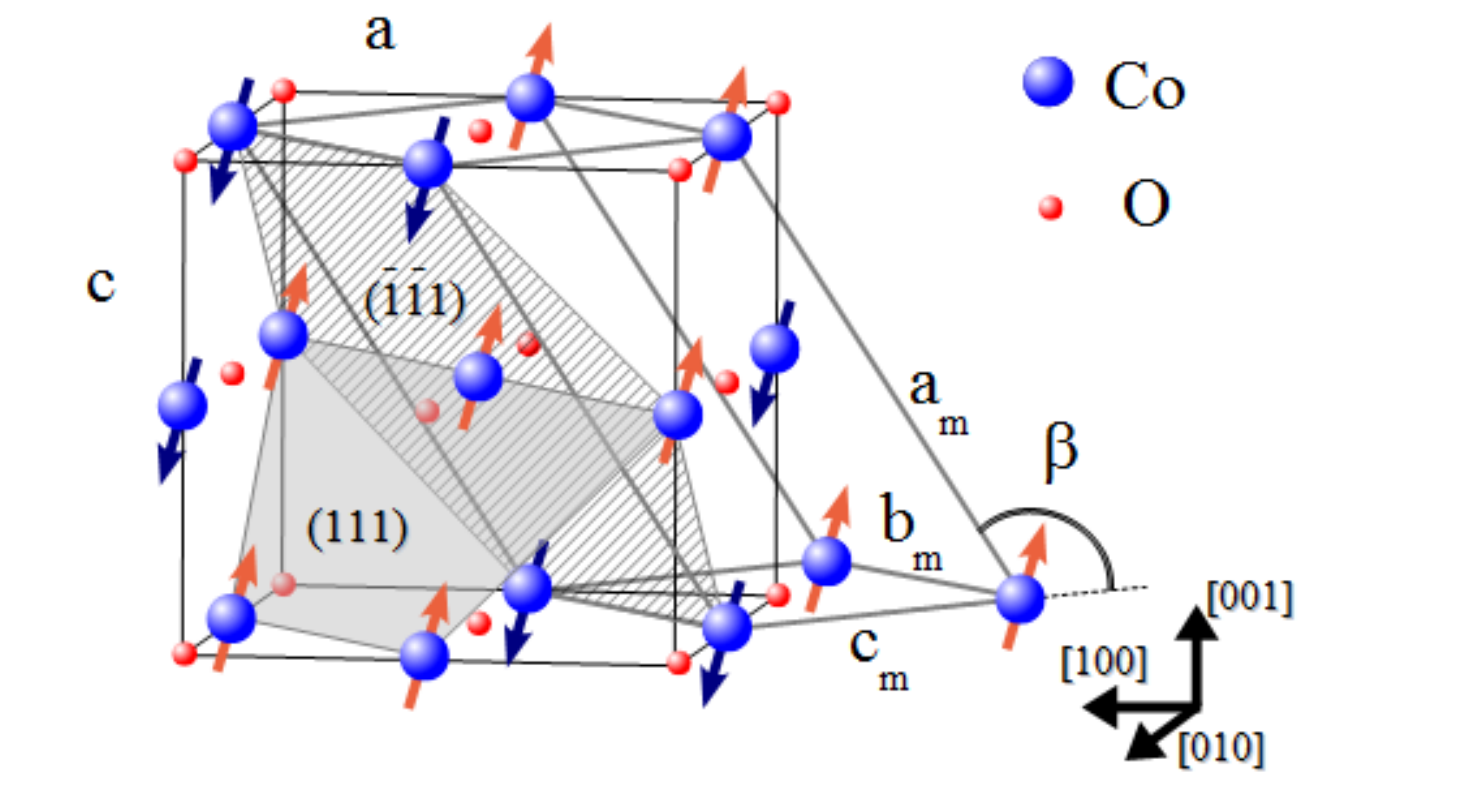}}
\caption{\label{fig:CoO} CoO rocksalt (${a}$, ${c}$) and monoclinic (${a_m}$, ${b_m}$, ${c_m}$, ${\beta}$) unit cell parameters and low temperature AFM spin structure \cite{Jauch2001PRB}. The gray half-hexagon indicates Co FM sheets on (111) planes, antiparallel along the [111] direction. The hatched hexagon indicates Co AFM sheets on ($\bar{1}$$\bar{1}$1) planes. }
\end{figure}

Ultrathin CoO films number among the most interesting AFM layers for spintronic devices. At room temperature (RT), bulk CoO paramagnetic phase crystallizes in the rocksalt structure where  pure Co and O  planes  alternate along the [111] axis (fig.\ref{fig:CoO}). It has a N\'eel temperature (${T}_{N}$) of 293 K and a magnetic moment  of 3.98 ${\mu}_{B}$  \cite{Roth1958PR, Jauch2001PRB}. The magnetic moment lies far above the 3 ${\mu}_{B}$ value, revealing  a  large orbital contribution. The strong interaction between spin and orbital magnetic moments through the spin-orbit coupling drives the magnetic anisotropy energy \cite{Schron2012PRB}. Below ${T}_{N}$, the AFM ordering develops concomitantly with a monoclinic distorted phase \cite{Jauch2001PRB}.
The AFM structure is described as a stacking of FM hexagonal sheets of high spin ${\mathrm{Co}}^{2+}$ ions coupled antiferromagnetically along the [111] direction.  The spin structure is found collinear, with the moments close to the [001]  axis of the rocksalt lattice (fig.\ref{fig:CoO}).
The concomitant changes in the structure and magnetic properties suggest that distortion and antiferromagnetism are linked by magnetostriction  \cite{Roth1958PR, Jauch2001PRB,Schron2012PRB}. This view is supported by soft x-ray absorption spectroscopy (XAS) experiments in thin CoO layers grown on different substrates, which revealed significant modifications in the magnitude and orientation of the magnetic moments induced by epitaxial strain \cite{Csiszar2005PRL}. 
A critical issue  for competitive CoO based devices is, however, the preservation of a significant EB effect up to temperatures as close as possible to RT. Nevertheless, so far, all  experimental studies in ultrathin (${<}$10 nm) CoO/FM double-layer systems report EB blocking temperatures (${T}_{B}$) smaller than ${T}_{N}$ \cite{vanderZaag2000PRL, Maat2001PRL,Tonnerre2008PRL}. The situation seems to be identical for FM layers with planar or with  perpendicular magnetic anisotropy (PMA).  

We report here on the exchange coupling and magnetic properties of an ultrathin CoO/PtFe double-layer. The growth by reactive molecular beam epitaxy of an ultrathin CoO on a Pt-terminated PtFe/Pt(001) surface leads to a hexagonal surface and monoclinic distorted CoO film at RT.  The strain-induced monoclinic distortion in the CoO film accounts for its in-plane spin orientation. The exchange coupling of the distorted CoO layer with the PtFe(001) layer brings forth a robust perpendicular EB shift, which is preserved up to the AFM ordering temperature. This finding provides a unique example where the blocking (${T}_{B}$) and ordering (${T}_{N}$) temperatures are identical and match the bulk N\'eel temperature. 

We have used ${in}$ ${situ}$ grazing incidence x-ray diffraction (GI-XRD) at the French CRG BM32 beamline at the European Synchrotron Radiation Facility (ESRF, France) to study and optimize the growth of ultrathin films on ultrahigh vacuum cleaned Pt(001) substrates \cite{Soares2012PRB}. 
Our Pt-terminated  PtFe layer was grown by thermal deposition of three monolayers (ML) of Fe on a clean Pt(001) substrate hold at 600 K, followed by 1 ML Pt deposition. This procedure gives rise to a 1.2 nm-thick ${L1}_{0}$  PtFe(001)  layer in coherent epitaxy on Pt(001). The ${L1}_{0}$ phase, formed by alternate Fe and Pt atomic planes along the c-axis of the tetragonal structure, provides an out-of-plane spin network with strong perpendicular magnetic anisotropy \cite{Okamoto2002PRB,Soares2011JAP}. 
The CoO layer was grown by reactive molecular beam epitaxy on the ultrathin PtFe(001) layer hold at 523 K.  
Owing to the high oxidation potential of Fe \cite{Regan2001PRB}, reactive CoO deposition on pure Fe oxidizes about 1-2 ML of Fe \cite{Regan2001PRB,Wu2010PRL, Abrudan2008PRB}. Our Pt-terminated high quality PtFe(001) layer shows a small  oxide contribution, likely related to Fe atoms dispersed within the CoO layer or from Fe-O bounds at the interface. We show  here below  that the magnetic properties are characteristic of metallic ${L1}_{0}$  PtFe and  are not affected by the small oxide contribution.
CoO thickness was chosen around 3 nm, close to the onset thickness for frozen AFM spins  \cite{Wu2010PRL}. The detailed growth procedure and x-ray diffraction study of CoO on both PtFe(001) and Pt(001) will be presented elsewhere \cite{Lamirand2013a}.  

\begin{figure}
\resizebox{1\columnwidth}{!}{\includegraphics{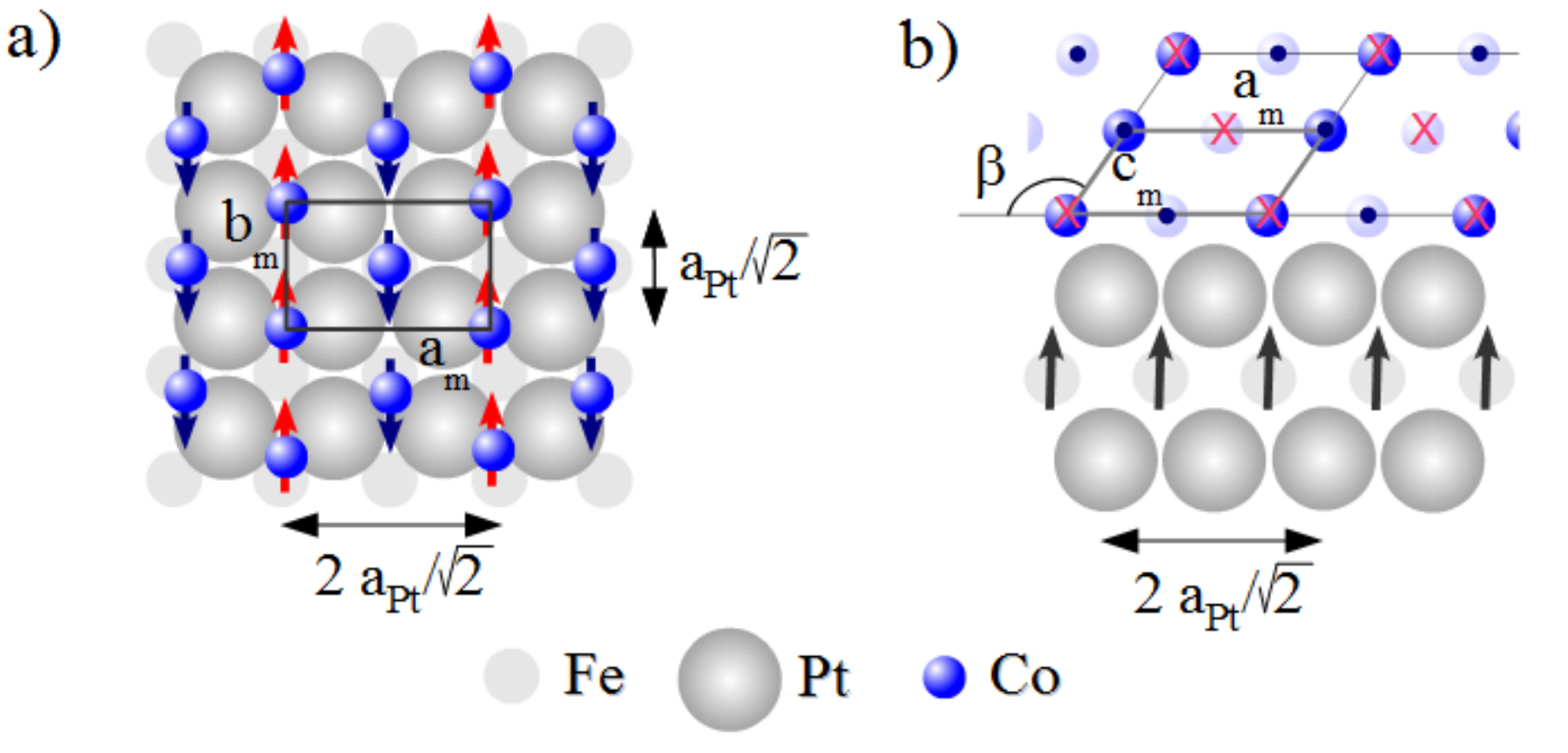}}
\caption{\label{fig:CoO_fig2}  Illustration of the hexagonal (111)-like CoO layer on PtFe/Pt(001). Spin orientation and relation between the CoO monoclinic (${a_m}$, ${b_m}$, ${c_m}$, ${\beta}$) and Pt (a${_{Pt}}$) parameters.  (a) Top view: Co AFM spin structure (fig.1) projected onto the surface, with spin axis along the [1${\bar1}$0] direction. (b) Side view:  Fe spins are perpendicular to the surface and the projected Co AFM spins point forward (${\bullet}$) or backward (${\times}$). }
\end{figure}

The GI-XRD analysis at RT shows that hexagonal Co atomic planes sit on the underlying Pt-terminated square network (fig.\ref{fig:CoO_fig2}-a). Such a hexagonal (111)-like CoO surface is not uncommon on substrates with similar lattice misfit between the oxide and substrate \cite{Mittendorfer2012PRL}. The 2D-rectangle network (${a_m}$, ${ b_m}$) corresponding to CoO(111) hexagonal planes does not exactly match the 2D-rectangle network (${2a}_{Pt}/\sqrt{2}$, ${a}_{Pt}/\sqrt{2}$) defined by the Pt underlayer. The misfits along  ${a_m}$ and ${b_m}$ axis ([112] and [1$\bar{1}$0] rocksalt directions) are  ${\epsilon_a}$=-6.1\% and ${\epsilon_b}$=+8.6\%, respectively, bringing about a slightly anisotropic stress. Consequently, the 3 nm-thick CoO layer is slightly compressed and develops a small monoclinic distortion (${\beta}{\neq}{\beta{_o}}$=125.264$^{\circ}$). About 8 nm-large well-crystallized domains are observed for the four orientations allowed by symmetry \cite{Lamirand2013a}. The RT monoclinic cell parameters obtained from GI-XRD analysis averaged over all domains are ${a_m}$=5.220(2)\AA${}$, ${b_m}$=3.005(1)\AA${}$, ${c_m}$=2.995(3)\AA${}$ and ${\beta}$=124.995(5)$^{\circ}$. In the slightly deformed tetragonal lattice, the parameters are ${a}$=4.243(3)\AA${}$ and ${c}$=4.272(3)\AA${}$.

%
Element-resolved spin orientation  of the  CoO/PtFe/Pt(001) system was  ${ex}$ ${situ}$ investigated  by XAS, using linear and circular magnetic dichroism  at Fe and Co  ${L}_{2,3}$ edges. XAS measurements under applied magnetic field were performed  at the PGM beamline of the Laborat\'orio Nacional de Luz S\'incrotron (LNLS, Brazil), with a spectral resolution of  E/${\Delta}$E=6000 and degrees of linear and circular polarizations close to 100\% and 80\%, respectively. The sample was allowed to rotate about a vertical axis, with the polar angle ${\theta}$  defined as the angle between the surface normal and the x-ray propagation. X-ray linear dichroism (XLD) was taken with linear horizontal polarization, as the difference between XAS recorded at  ${\theta}$ and  ${\theta}$=0$^{\circ}$. X-ray magnetic circular dichroism (XMCD) was taken as difference between right and left circular polarizations at  ${\theta}$=0$^{\circ}$. All  spectra were collected  using total electron yield, corrected for electron yield saturation effects \cite{Regan2001PRB} and normalized far from ${L}_{2,3}$  edges. The XAS  study has  been complemented by polar magneto-optic Kerr effect (MOKE) measurements. All magnetic measurements were  performed after  a field cooling from 350 K down to 5 K under an applied magnetic field of +5 kOe along the normal to the sample surface.  

\begin{figure}
\resizebox{1\columnwidth}{!}{\includegraphics{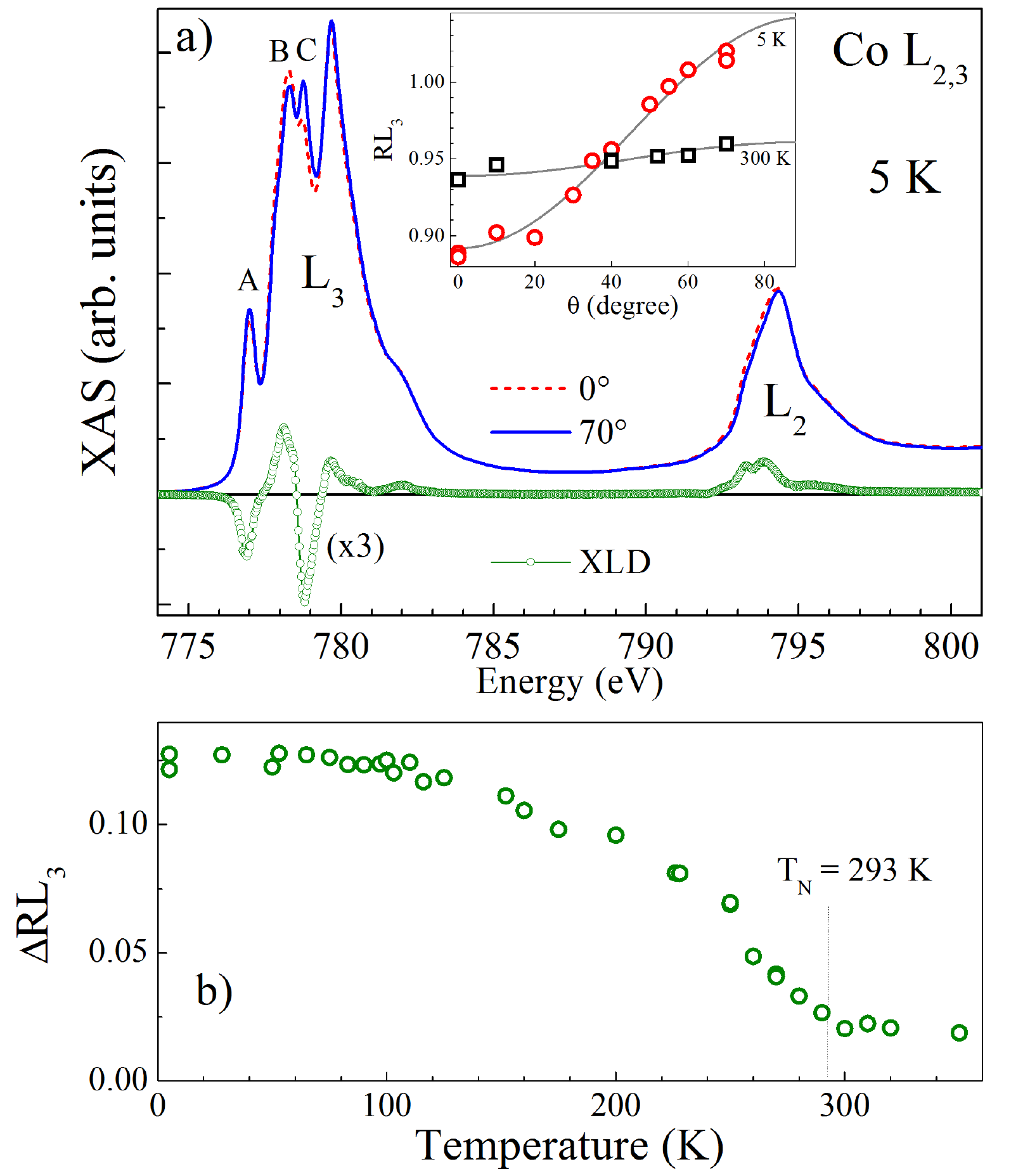}}
\caption{\label{fig:XMLD} (a) Co ${L}_{2,3}$ XAS spectra of a 3 nm-thick CoO layer on PtFe/Pt(001) at 5 K after cooling in a +5kOe magnetic field. Linear polarization parallel to the surface (${\theta}$=0$^{\circ}$, dashed (red) line) and  towards the surface normal (${\theta}$=70$^{\circ}$, solid (blue) line). The XLD (dot (green) line) is the difference between them. The angle-dependent anisotropy, defined by the C over B peak contrast, is shown in the inset. (b) Temperature dependence of the anisotropy.  }
\end{figure}

Figure \ref{fig:XMLD}-a shows the Co ${L}_{2,3}$  XAS spectra at 5 K for ${\theta}$=0$^{\circ}$  and ${\theta}$=70$^{\circ}$. The difference between them gives a clear XLD signal, which  essentially  measures the charge anisotropy associated to  both the local crystal field and  the local exchange field through the spin-orbit coupling \cite{vanderLaan2008PRB, Wu2010PRL, Csiszar2005PRL}. The latter contribution is known as x-ray magnetic linear dichroism (XMLD).
Following  Wu and coworkers  \cite{Wu2010PRL} we used the intensity ratio ${RL}_{3}$  between the peaks at 778.74 eV (C) and 778.26 eV (B) as a measure of the overall anisotropy. 
The XMLD contribution to  ${RL}_{3}$ is maximum at the angle where the polarization vector is perpendicular to the Co magnetic moments \cite{vanderLaan2008PRB, Wu2010PRL}. 
A ${cos}^{2}{\theta}$ fit of ${RL}_{3}$  for the low temperature measurements  (inset fig.\ref{fig:XMLD}-a) has its minimum when the polarization is parallel to the surface (${\theta}$=0$^{\circ}$) and its maximum at about  ${\theta}$=90$^{\circ}$, within an accuracy of a few degrees. We can then conclude that the Co spin axis is essentially parallel to the surface.  As the Fe spin axis is perpendicular to the surface, the coupling between Co and Fe spins through the Pt interface layer is at 90$^{\circ}$ (fig.\ref{fig:CoO_fig2}). 
Further information on the spin orientation within the film plane can be sought from the  rich manifold peak structure, including the ${L}_{3}$ and ${L}_{2}$  Co edges. Atomic multiplet calculations performed by van der Laan and coworkers \cite{vanderLaan2008PRB} show that Co magnetization axis with respect to the crystalline axis can be differentiated from the relative variations of ${L}_{3}$ and ${L}_{2}$ features and from those of ${L}_{3}$  A (at 777.0 eV) and C peaks (fig.5 in Ref.\cite{vanderLaan2008PRB}). The XLD signal in figure \ref{fig:XMLD}-a matches the situation where Co spins are along the ${b_m}$ axis ([1${\bar1}$0] direction) (fig.\ref{fig:CoO_fig2}). 

\begin{figure}
\resizebox{1\columnwidth}{!}{\includegraphics{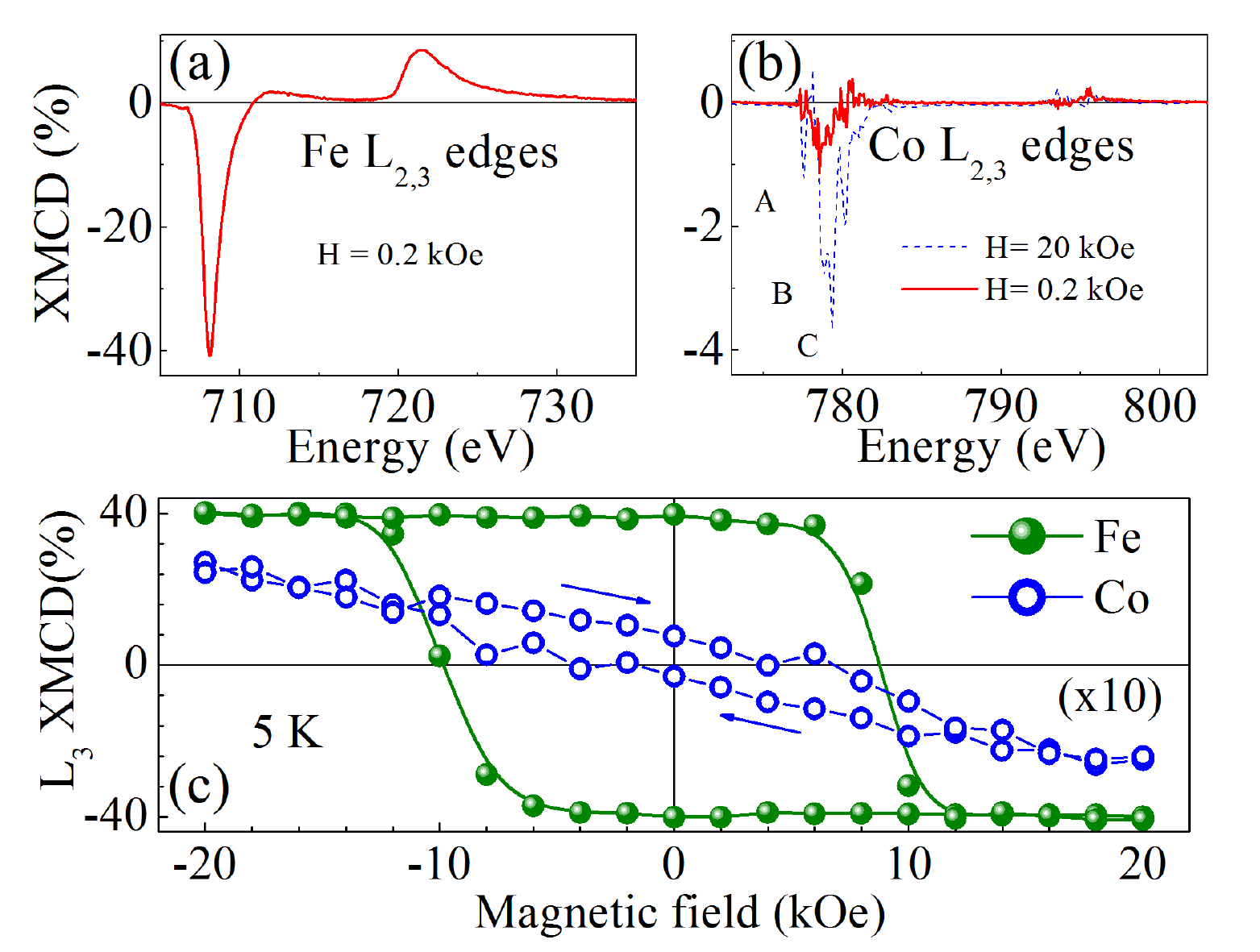}}
\caption{\label{fig:XMCD}  (a) Fe ${L}_{2,3}$ and (b) Co ${L}_{2,3}$ XMCD: solid (red) lines are close to remanence; dashed (blue) line is at 20 kOe. Note the factor of ten between scales. (c) Element selective hysteresis loops at Fe and Co ${L}_{3}$ edges.}
\end{figure}

As a small out-of-plane contribution of  interface Co spin cannot be ruled out from the only  XMLD, we went on performing XMCD measurements at Fe and Co ${L}_{2,3}$  edges (fig.\ref{fig:XMCD}).  Element-selective hysteresis loops were drawn  by reporting for each value of the applied perpendicular magnetic field, the maximum amplitude of the XMCD at the Fe and Co ${L}_{3}$  edges. Fe ${L}_{2,3}$ XMCD at 5 K (fig.\ref{fig:XMCD}-a) has a metallic signature and a maximum amplitude of about 40\% at the ${L}_{3}$ edge. The Fe hysteresis loop is shifted towards negative values and yields a magnetization at zero field (remanence) close to the saturation magnetization (fig.\ref{fig:XMCD}-c). Such almost 100\% remanence indicates the PMA character of the PtFe layer, which has been confirmed by a hysteresis loop measured at ${\theta}$=70$^{\circ}$ (not shown).
At the Co ${L}_{3}$ edge, we observe a weak XMCD signal due to the Co spin component not compensated by AFM interactions (fig.\ref{fig:XMCD}-b).  It shows the CoO characteristic  multiplet features \cite{vanderLaan2008PRB, Abrudan2008PRB}. 
The  maximum amplitude of this XMCD signal is roughly proportional to the applied magnetic field but  shows a weak hysteresis opening with a remaining contribution of about 0.5(3)\% close to remanence and coinciding with the Fe hysteresis loop  (fig.\ref{fig:XMCD}-c). Two contributions to the field dependent Co XMCD should then be considered. The  linear contribution is a bulk-like effect,  arising from the coupling of the whole set of Co spins in the CoO layer to the  external magnetic field. On the other hand, the weak hysteresis, following the  Fe  hysteresis loop, results from an interface exchange coupling with Fe. This small contribution originates from an uncompensated Co spin component perpendicular to the surface. 

We turn now to the temperature dependent magnetic properties. 
In the ${RL}_{3}$ ratio (fig.\ref{fig:XMLD}-a), the magnetic and structural contributions to the XLD signal are mixed up by the local tetragonal crystal field. This explains the  small residual  anisotropy observed at 300 K  (fig.\ref{fig:XMLD}-a, inset),  where no magnetic contribution is expected. Magnetic and non magnetic contributions can be disentangled by a full temperature dependence study of the anisotropy amplitude,  experimentally defined  as ${\Delta}$${RL}_{3}$=${RL}_{3}$(70$^{\circ}$)-${RL}_{3}$(0$^{\circ}$).  ${\Delta}$${RL}_{3}$  decreases following a Brillouin-like function up to  ${T}_{N}$  $\approx$ 293 K  and then stabilizes (fig.\ref{fig:XMLD}-b). This unambiguously confirms that the AFM order is preserved up to about 293 K. It also proves that the N\'eel temperature of the CoO film is very close to that of the bulk CoO crystal \cite{Jauch2001PRB}. Above  ${T}_{N}$  only the non magnetic crystal field contribution  to the anisotropy still remains.

\begin{figure}
\resizebox{1\columnwidth}{!}{\includegraphics{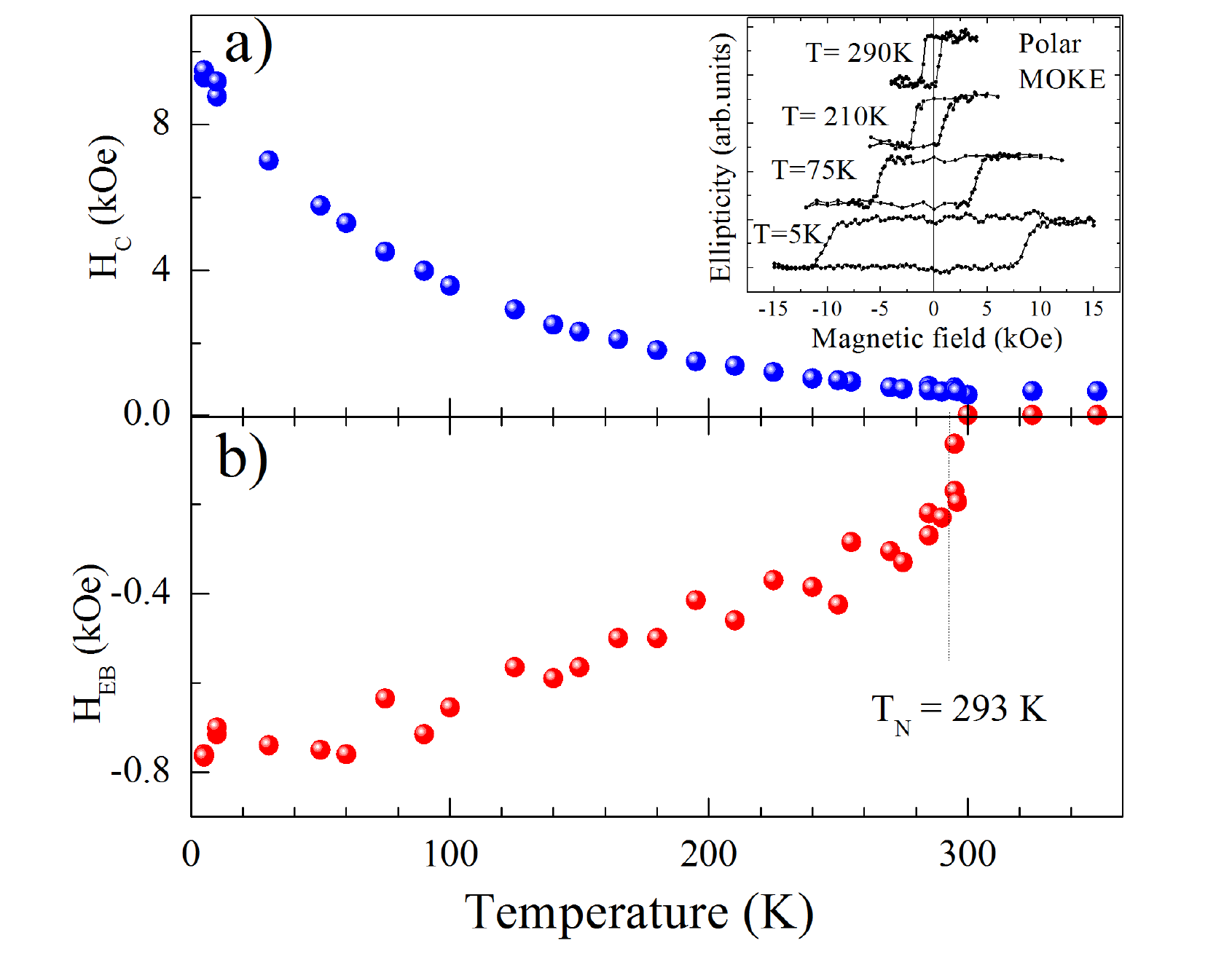}}
\caption{\label{fig:MOKE}Temperature dependence of the (a) coercive field and (b) exchange bias after cooling the sample in a +5 kOe perpendicular magnetic field.  Polar MOKE hysteresis loops at a few selected temperatures are shown in the inset.}
\end{figure}

We  will  now focus on  the temperature dependence of the coercive field (${H_C}$) and exchange bias shift (${H}_{EB}$)  measured by polar MOKE (fig.\ref{fig:MOKE}). The hysteresis loop shows 100\% remanence from the lowest to the highest temperature (fig.\ref{fig:MOKE}, inset), confirming that the PMA of the ultrathin PtFe layer is preserved. The loops also show a perpendicular exchange bias  shift  up to the nominal bulk N\'eel temperature (fig.\ref{fig:MOKE}-b), asserting the steadfastness of the CoO/PtFe exchange interfacial coupling and also of the AFM order of the CoO layer. The exchange bias shift is about ${H}_{EB}$=-0.75 kOe at 5 K.


In low temperature bulk CoO, the monoclinic distortion is essentially driven the Jahn-Teller effect due to the partial filling of the ${\mathrm{Co}}^{2+}$ ${t}_{2g}$ orbitals \cite{Schron2012PRB}. It can be seen as the result of a main tetragonal plus a small trigonal distortion. The magnetic dipole-dipole interaction tends to align spins in the (111) plane, while crystal field energy, arising from the compressive (${c/a<}$1) tetragonal deformation, favors the [001] direction \cite{Roth1958PR}. As  a consequence, the CoO spin structure is collinear with the spin axis making a small angle with the [001] direction of the rocksalt lattice \cite{Roth1958PR,Jauch2001PRB} (fig.1). 
In epitaxial thin films the strain drives the anisotropy. Csiszar and coworkers \cite{Csiszar2005PRL} have shown that an ultrathin CoO layer sandwiched by MnO layers on Ag(001) shows an out-of-plane magnetization axis along [001], while  in direct epitaxy the Ag(001) substrate  it shows an in-plane magnetization axis orthogonal to the  [001] direction. The main structural difference between the two cases  lies in the CoO${(001)}$ epitaxial strain, which is tensile on MnO${(001)}$ and slightly compressive on Ag${(001)}$, generating respectively a compressive (${c/a<}$1) and a slightly extensive (${c/a>}$1) tetragonal deformation. 
The slightly anisotropic strain imposed by the PtFe/Pt(001) surface on the CoO layer leads to a monoclinic distorted lattice that resembles that of bulk CoO at low temperatures. However, while the tetragonal deformation is  compressive  in  the bulk (${c/a}$=0.988),  it is  extensive in the film (${c/a}$=1.008). In this particular situation, the dipole-dipole magnetic energy is minimized when the FM Co spins are within the (111) plane and parallel to the [1${\bar1}$0] direction \cite{Finazzi2003PRB}.

It is noteworthy that,  owing  to the  monoclinic distortion, the hexagonal (111) plane perpendicular to the trigonal distortion is no longer equivalent to the other hexagonal planes. From strict structural considerations, we can identify the hexagonal plane sitting on the PtFe(001) surface as the one parallel to the (${a_m, b_m}$) plane and not the one perpendicular to the trigonal elongation (fig.\ref{fig:CoO}, hatched (${\bar1}$${\bar1}$1) and gray (111) hexagons, respectively). It is then expected the Co sheets parallel to the surface to be those containing fully compensated spins (fig.\ref{fig:CoO_fig2}). In this plane rows of Co spins are coupled ferromagnetically along ${b_m}$ ([1${\bar1}$0] direction) and antiferromagnetically along ${a_m}$ ([112] direction).  Such AFM configuration resembles the model predicted by DFT calculations for a single CoO overlayer on Ir(001) \cite{Mittendorfer2012PRL}.
The Co spin orientation (fig.\ref{fig:CoO_fig2}) deduced from our XLD analysis is fully consistent with the sequence of alternate  FM Co(111) planes,  but the spin axis here is along the [1${\bar1}$0] direction and does not contain any component out of the hexagonal surface.

%
Many experimental studies report that to reach blocking (${T}_{B}$) temperatures close to  ${T}_{N}$, CoO thickness should be at least about 10 nm  \cite{Berkowitz1999JMMM, vanderZaag2000PRL, Ambrose1996PRL,   MolinaRuiz2011PRB}. In most cases, the blocking temperature measured from the onset of the exchange bias shift is smaller than the expected  ${T}_{N}$.  Films with thickness around 3-5 nm display ${T}_{B}$ typically around 200-240 K.
In constrast, our CoO layer sustains an EB shift up to ${T}_{N}{\approx}$ 293 K. This exceptional behavior must be related to the good crystalline quality and to the stable spin configuration at the interface. It demonstrates that AFM order as in the bulk may be established in CoO films as thin as 3 nm and that the thickness effect, which reduces the ordering temperature, is not an intrinsic property.

CoO layers may couple with FM layers showing in-plane \cite{vanderZaag2000PRL, Wu2010PRL, Abrudan2008PRB} or out-of-plane \cite{Maat2001PRL,Tonnerre2008PRL} anisotropy. Exchange coupling properties are largely determined by the direction and strength of the anisotropy in the FM and in the AFM layers.
The XMCD study reveals that there is at the interface a weak uncompensated Co spin component perpendicular to the surface. However,  Co spins are essentially aligned in-plane. The coupling between interfacial Co and Fe spins is then at 90$^{\circ}$ (fig.\ref{fig:CoO_fig2}-b). A similar 90$^{\circ}$ coupling is not unusual and has been reported for in-plane anisotropy systems as CoO/Fe on Ag(001) \cite{Wu2010PRL}. Such an orthogonal coupling minimizes the energy for a fully compensated AFM interfacial spin configuration interacting with the exchange field of the FM layer.
In addition, we should remind that  in the PtFe layer the high  magnetic anisotropy relies on the strong spin-orbit coupling of the Pt site and hybridization between Fe 3d and Pt 5d states \cite{Nakajima1998PRL}. Exchange coupling of Co and Fe moments through Pt 5d states at the interface likely contributes to the preservation of the EB shift up to the AFM phase transition.

To summarize, the growth by reactive molecular beam epitaxy of a 3nm-thick CoO layer on a Pt(001)-terminated PtFe(001) surface gives rise to a hexagonal CoO(111)-like surface, which develops into a monoclinic distorted film at RT. Using polarization dependent XAS at Co and Fe ${L}_{2,3}$ edges, we have given a complete description of the orientation of the Co and Fe magnetic moments. We have shown that the coupling of such a distorted CoO hexagonal layer with PMA PtFe(001) brings forth a very robust perpendicular exchange bias shift preserved  up to the antiferromagnetic ordering temperature of 293 K. This is a unique example where the blocking and N\'eel temperatures for an  ultrathin  CoO layer are identical and match the bulk N\'eel temperature.  Such exceptional behavior shares a close relationship with the strain-induced distortion of the oxide layer.  Our outcome demonstrates that the thickness effect on ${T_N}$, which reduces the ordering temperature, and reduction of blocking temperature ( ${T_B}$) are not intrinsic properties of these double-layers.

\begin{acknowledgments}
Beamtime is acknowledged at the PGM/LNLS beamline and at the French CRG BM32/ESRF beamline. We are greatfull to the PGM beamline staff for the skillful technical assistance and C. Vergnaud for support in the MOKE measurements.
\end{acknowledgments}

\providecommand{\noopsort}[1]{}\providecommand{\singleletter}[1]{#1}%

\end{document}